\documentclass[12pt]{article}
\usepackage[dvips]{graphicx}
\begin{document}
\title{Two-dimensional loosely and tightly bound solitons in optical
lattices and inverted traps}
\author{H Sakaguchi$^{\dag}$ and B A Malomed$^{\ddag}$\\
\dag\ Department of Applied Science for Electronics and Materials,\\
Interdisciplinary Graduate School of Engineering Sciences,\\
Kyushu University, Kasuga, Fukuoka 816-8580, Japan\\
\ddag\ Department of Interdisciplinary Studies,Faculty of Engineering,\\
Tel Aviv University, Tel Aviv 69978, Israel}
\maketitle
\begin{abstract}
We study the dynamics of nonlinear localized excitations (\textquotedblleft
solitons") in two-dimensional (2D) Bose-Einstein condensates (BECs) with
repulsive interactions, loaded into an optical lattice (OL), which is
combined with an external parabolic potential. First, we demonstrate
analytically that a broad (\textquotedblleft loosely bound\textquotedblright
, LB) soliton state, based on a 2D Bloch function near the edge of the
Brillouin zone (BZ), has a negative effective mass (while the mass of a
localized state is positive near the BZ center). The negative-mass soliton
cannot be held by the usual trap, but it is safely confined by an inverted
parabolic potential (\textit{\ anti-trap}). Direct simulations demonstrate
that the LB solitons (including the ones with intrinsic vorticity) are
stable and can freely move on top of the OL. The frequency of elliptic
motion of the LB-soliton's center in the anti-trapping potential is very
close to the analytical prediction which treats the solition as a
quasi-particle. In addition, the LB soliton of the vortex type features real
rotation around its center. We also find an abrupt transition, which occurs
with the increase of the number of atoms, from the negative-mass LB states
to tightly bound (TB) solitons. An estimate demonstrates that, for the
zero-vorticity states, the transition occurs when the number of atoms
attains a critical number $N_{\mathrm{cr}}\sim 10^{3}$, while for the vortex
the transition takes place at $N_{\mathrm{cr}}\sim 5\cdot 10^{3}$ atoms. The
positive-mass LB states constructed near the BZ center (including vortices)
can move freely too. The effects predicted for BECs also apply to optical
spatial solitons in bulk photonic crystals.
\end{abstract}

\section{\noindent Introduction}

Periodic potentials, usually called optical lattices (OL), which are induced
by interference of laser beams illuminating a Bose-Einstein condensate
(BEC), are widely used in experimental studies of BECs. Not only
one-dimensional (1D) OLs, but also their multidimensional counterparts are
available in the experiment, see Refs. \cite{OLexperiment} and references
therein.

An important type of nonlinear excitations that BECs may support are bright
solitons (in this paper, following the trend commonly adopted in the current
literature, we apply the term \textquotedblleft soliton" to robust localized
objects, without implying integrability of the underlying mathematical
model). In an effectively 1D (\textquotedblleft
cigar-shaped\textquotedblright ) condensate, solitons have been observed in
the experiment \cite{soliton_exper}. Creation of solitons in a BEC confined
in the OL has not been reported yet, although splitting of wave packets in a
$^{87}$Rb condensate, loaded in a cylindrical trap with an superimposed 1D
OL potential, which was reported recently \cite{soliton_transient}, might be
a manifestation of transient soliton dynamics.

Parallel to the experiments, many theoretical predictions for BEC
solitons in OLs have been reported. In particular, attention was
attracted to a possibility to create \textit{gap solitons}, that
are expected as a result of the interplay of the OL potential and
\emph{repulsive} interaction between atoms in the BEC
\cite{SalernoJPhysB,othergapsol}. While only 1D solitons have been
thus far observed in BECs, it was predicted that OLs may support
multidimensional solitons of the gap type as well \cite
{SalernoJPhysB}. A prediction which is especially relevant to
feasible experiments is that OLs can also stabilize
multidimensional solitons in BECs with attraction between atoms
(because of the possibility of collapse, such solitons are
unstable without a support potential) \cite{BMS}. Moreover, 2D
bright solitons carrying intrinsic vorticity can be stabilized too
by the OL in the 2D case \cite{BMS,YangMuss}. Various other
properties of OL-supported solitons were studied too
\cite{otherOLsol}.

Similar localized objects are expected in a different physical
setting, in the form of 2D optical spatial solitons in nonlinear
photonic crystals (NPCs) \cite{PhotCryst,BMS,YangMuss}. In optical
media, an effective lattice potential is induced by periodic
modulation of the local refractive index in the transverse
direction(s) \cite{Wang,BMS,YangMuss}. An alternative is to use a
\textit{virtual lattice}, induced in a photorefractive medium by a
transverse array of strong coherent laser beams; the latter
arrangement has made it possible to observe 2D spatial solitons in
a real experiment \cite {Moti}, as well as their
vorticity-carrying counterparts \cite{photrefvortex} .

Most theoretical studies of solitons in the above-mentioned models
did not address their motion. In a recent work \cite{rf:5}, we
considered in detail the motion and collisions of 1D solitons in
the OL-equipped model in both cases of the attractive and
repulsive cubic nonlinearity. In that work, the OL potential was
combined with an external parabolic (harmonic) trapping potential,
which corresponds to the real situation in any BEC experiment.
Direct simulations of the corresponding effective 1D
Gross-Pitaevskii (GP) equation had demonstrated that, in either
case (self-focusing or self-defocusing nonlinearity), the soliton
travels freely through the lattice if its norm (number of atoms)
is below a certain critical value; above this value, it gets
trapped by the OL. Depending on parameters, collisions between
moving solitons are either elastic, or lead to their fusion. The
most intriguing result concerns the motion of the soliton in the
parabolic trap: in the case of attraction, the soliton behaves, as
it may be naturally expected, as a quasi-particle in the
harmonic-oscillator potential. However, in the case of repulsion,
the soliton of the gap type is expelled by the parabolic trap,
while it is stably confined by an \textit{ anti-trap} (repulsive
parabolic potential). This result, which calls for direct
experimental verification, was supported by analytical
considerations which demonstrate that the effective mass of the
gap soliton, regarded as a quasi-particle, is \emph{negative} (in
the case of the usual soliton, corresponding to the attractive
nonlinearity, the effective mass is positive).

An issue of straightforward interest is to extend these results to the 2D
case, which is quite relevant, as experiments with 2D solitons in BECs seem
possible right now. Besides that, the problem is also of direct relevance
for steering spatial optical soliton in NPC media. In \ this work, we focus
on the 2D BEC model with repulsive interactions, as this case is most
relevant to the experiment, and produces most interesting physical results.
First, we demonstrate that broad localized states, which we call loosely
bound (LB) ones, may be represented as a product of a linear Bloch function,
defined by the OL, which is taken close to the edge to the corresponding
Brillouin zone (BZ), and a slowly varying envelope function, which obeys an
asymptotic nonlinear equation that does not contain OL. Such a
representation makes it possible to demonstrate analytically that the
effective mass of the LB state is negative. This way, we find not only
fundamental LB states, but also ones carrying vorticity, whose mass is
negative too; the existence of 2D localized vortices in the repulsive model
has not been reported before. In direct simulations, the\ LB vortex features
actual rotation around its center. Simulations confirm the stability of the
localized LB states, and demonstrate that they can move freely through the
OL. In particular, both fundamental and vortex LB states can be readily set
in elliptic circular motion in the anti-trapping potential, the frequency of
the circulations being very accurately predicted by the asymptotic equation.

Direct simulations (as well as the asymptotic equation) demonstrate an
abrupt transition (without hysteresis) from the LB states to much stronger
localized tight-bound (TB) ones with the increase of the norm (number of
atoms). The TB states are nothing else but the 2D gap solitons, that were
first found in Ref. \cite{SalernoJPhysB}. The LB-TB transition also takes
place for the vortices, thus revealing the existence of new TB vortex
solitons of the gap type. It is additionally found that the TB vortices
undergo an intrinsic transition to a more symmetric shape with subsequent
increase of the norm. Unlike their LB counterparts, the TB solitons cannot
move, being strongly pinned by the OL.

We also construct LB states, starting from the linear Bloch function close
to the BZ center. In that case, the effective mass is positive, no
transition to TB solitons occurs (as they do not exist), and the LB states,
both fundamental and vortex ones, are mobile, which is accurately described
by the corresponding asymptotic equation.

The rest of the paper is organized as follows. The analysis based on the
derivation of the asymptotic equations for the LB states, with both negative
and positive mass, is presented in section 2. Numerical results for static
solitons (however, including rotation of the vortex LB states), such as the
LB-TB transitions, are summarized in section 3. In section 4, we present
numerical and analytical results concerning motion of the LB states, for the
cases of the negative and positive mass. The paper is concluded by section 5.

\section{Formulation of the model and the analytical approximation}

\subsection{The Gross-Pitaevskii equations}

The mean-field description of the BEC dynamics is based on the GP equation
for the mean-field wave function $\psi $ in three dimensions \cite{Pit},
\begin{equation}
i\hbar \frac{\partial \psi }{\partial t}=\left[ -\frac{\hbar ^{2}}{2m}\nabla
^{2}+U\left( \mathbf{r}\right) +G|\psi |^{2}\right] \psi ,  \label{GP3D}
\end{equation}
where $m$ is the atomic mass, $G\equiv 4\pi \hbar ^{2}a/m$, $a$ is
the $s$ -wave scattering length, $U$ is the potential applied to
the condensate, and the number of atoms is $\int |\psi
|^{2}d\mathbf{r}$. As is well known \cite
{SalernoJPhysB,othergapsol,Isaac}, in the case of a pancake-shaped
configuration (when the frequency characterizing the confining
potential in the transverse direction is much larger than the
confining frequencies corresponding to the perpendicular
directions), Eq. (\ref{GP3D}) can be reduced to a 2D equation.
After appropriate rescalings, such as $t^{\prime}=\hbar
t/(m\lambda^2), x^{\prime}=x/\lambda,y^{\prime}=y/\lambda$ and $\phi=G^{1/2}\psi$,
where $\lambda$ is the period of the spatially periodic potential,
it takes the form
\begin{equation}
i\phi _{t}=-\frac{1}{2}\nabla ^{2}\phi +\left\{ |\phi
|^{2}-\varepsilon \left[ \cos (2\pi x)+\cos (2\pi y)\right]
+\frac{1}{2}B(x^{2}+y^{2})\right\} \phi ,  \label{GP}
\end{equation}
where  $\phi $ is the renormalized wave function, the sign $^{\prime}$
denoting the rescaled coordinates $x^{\prime},y^{\prime},t^{\prime}$ is omitted,
the kinetic-energy operator
$\nabla ^{2}$ acts on the coordinates $x$ and $y$, the repulsive interaction
between atoms is assumed, the OL period is normalized to be $1$ in both
directions (we consider the simplest case of the square lattice, although
more sophisticated patterns, such as triangular, hexagonal, and
quasi-periodic ones, may be interesting too), while $\varepsilon $ and $B$
are amplitudes of the, respectively, OL and parabolic potential (in
simulations, we will fix $\varepsilon =10$ and $B=-0.005$, or $\varepsilon
=5 $ and $B=+0.005$, negative $B$ implying the anti-trapping potential).
We will show the numerical results with the dimensionless equation (2). The unit length corresponds to the spatial period $\lambda$ of the optical lattice, and the unit time corresponds to $m\lambda^2/\hbar\sim 1.3\times 10^{-3}$s for $\lambda\sim 10^{-6}$m and $m(^{87}{\rm Rb})\sim 1.37\times 10^{-25}$kg.

If $B=0$ and the nonlinear term is neglected, Eq. (\ref{GP}) becomes the
ordinary 2D linear Schr\"{o}dinger equation with the periodic potential,
which has Bloch-state solutions. The Bloch function $F(x,y)$ obeys the
equation
\begin{equation}
\mu F(x,y)=\frac{1}{2}\nabla ^{2}F+\varepsilon \left[ \cos (2\pi x)+\cos
(2\pi y)\right] F,  \label{F}
\end{equation}
where $\mu $ is the chemical potential (or simply energy, in the
case of a single atom), and the solution is quasi-periodic, i.e.,
$F(x,y)=\exp \left( ik_{x}x+ik_{y}y\right) ~f(x,y)$, where
$k_{x,y}$ are two wavenumbers, and the function $f(x,y)$ is
periodic, such that $ f(x+1,y)=f(x,y),f(x,y+1)=f(x,y) $. The
Bloch-wave solution determines the energy-band structure, $\mu
=\mu (k_{x},k_{y})$, the point $ (k_{x},k_{y})=(\pi ,\pi )$ being
the edge of the BZ (Brillouin zone).

Following the lines of the analytical approach to the 1D case developed in
Ref. \cite{rf:5}, an approximate solution to the full time-dependent GP
equation (\ref{GP}) for a small-amplitude broad localized state
(\textquotedblleft soliton") may be sought for, in the lowest approximation,
as
\begin{equation}
\phi (x,y,t)=e^{-i\mu t}F(x,y)\Phi (x,y,t),  \label{phi}
\end{equation}
where $F(x,y)$ is the Bloch function defined above, and $\Phi (x,y,t)$ is an
envelope function which varies slowly (as a function of $x$ and $y$) in
comparison with $F(x,y)$. This ansatz actually assumes that the nonlinearity
and parabolic potential in Eq. (\ref{GP}) are small perturbations in
comparison with the OL terms. The most typical gap soliton corresponds to a
solution which approaches the linear Bloch function with $k_{x}=k_{y}=\pi $
when its norm (the number of atoms), $N\equiv \int_{-\infty }^{+\infty
}|\phi |^{2}dxdy$, tends to zero (cf. a similar situation in the 1D case
\cite{rf:5}).

In the general case, the slowly varying amplitude $\Phi $ obeys an
asymptotic NLS equation, that can be derived by substituting the
ansatz (\ref {phi}) into Eq. (\ref{GP}) \ and applying a known
averaging procedure \cite {SalernoJPhysB}:
\begin{equation}
i\frac{\partial \Phi }{\partial t}=-\frac{1}{2}M_{\mathrm{eff}}^{-1}\nabla
^{2}\Phi +g|\Phi |^{2}\Phi +\frac{1}{2}B(x^{2}+y^{2})\Phi .  \label{asympt}
\end{equation}
Here, the effective mass $M_{\mathrm{eff}}$ for linear excitations is
determined from the curvature of the energy-band structure of the linear
Bloch states, $M_{\mathrm{eff}}^{-1}=(1/2)\left( \partial ^{2}/\partial
k_{x}^{2}+\partial ^{2}/\partial k_{y}^{2}\right) \mu $, and
\begin{equation}
g\equiv
\frac{\int_{0}^{1}\int_{0}^{1}|F(x,y)|^{4}dxdy}{\int_{0}^{1}
\int_{0}^{1}|F(x,y)|^{2}dxdy}~.  \label{g}
\end{equation}
The energy functional corresponding to the simplified GP equation
(\ref {asympt}) is
\begin{equation}
E=\frac{1}{2}\int \int dxdy\left[ M_{\mathrm{eff}}^{-1}\left\vert \nabla
\Phi \right\vert ^{2}+g\left\vert \Phi \right\vert ^{4}+B\left(
x^{2}+y^{2}\right) \left\vert \Phi \right\vert ^{2}\right] .  \label{E}
\end{equation}

Stationary solutions to the asymptotic equation (\ref{asympt}) can be looked
for in an obvious form,
\begin{equation}
\Phi =f(r)\exp \left[ i\left( m\theta -\mu t\right) \right] ,  \label{Phi}
\end{equation}
where $r$ and $\theta $ are polar coordinates in the plane $\left(
x,y\right) $, $m=0,1,2,...$ is a possible vorticity of the solution, and the
amplitude $f(r)$ obeys an equation
\begin{equation}
f^{\prime \prime }+r^{-1}f^{\prime }-m^{2}r^{-2}=-2M_{\mathrm{eff}}\left[
\mu -\left( B/2\right) r^{2}-gf^{2}\right] f  \label{eigen}
\end{equation}
[in the general case -- in particular, in the case of the full GP
equation ( \ref{GP}) which is not isotropic -- the vorticity of a
stationary solution may be  defined as $\Delta \theta /\left( 2\pi
\right) $, where $\Delta \theta $ is the total change of the
solution's phase along a closed path around the center]. The
linear limit of Eq. (\ref{eigen}) is the linear Schr\"{o}dinger
equation for the 2D harmonic oscillator. For fixed integer $m$,
the ground-state solution to the latter equation is (in the
present notation)
\[
f(r)=r^{m}\exp (-\alpha ^{2}r^{2}/2),
\]
where $\alpha ^{2}=\sqrt{BM_{\mathrm{eff}}}$, with $\mu
=(m+1)\omega $, and $ \omega =\sqrt{B/M_{\mathrm{eff}}}$.

\subsection{Solutions near the edge and center of the Brillouin zone}

Near the BZ edge, i.e., close to the point $\left( k_{x},k_{y}\right)
=\left( \pi ,\pi \right) $, the linear Bloch function can be approximated by
a combination of four harmonics:
\[
F(x,y)=c_{1}\exp (ik_{x}x+ik_{y}y)+c_{2}\exp \left( i(k_{x}-2\pi
)x+ik_{y}y\right)
\]
\begin{equation}
\;\;\;+c_{3}\exp \left( ik_{x}x+i(k_{y}-2\pi )y\right) +c_{4}\exp \left(
i(k_{x}-2\pi )x+i(k_{y}-2\pi )y\right) .  \label{Bloch}
\end{equation}
The substitution of this into Eq.~(\ref{F}) and a usual truncation procedure
(neglecting higher-order spatial harmonics) lead to an expression $\mu
(k_{x},k_{y})=\mu _{0}^{\mathrm{(edge)}}+\left[ (k_{x}-\pi )^{2}+(k_{y}-\pi
)^{2}\right] /(2M_{\mathrm{eff}}^{\mathrm{(edge)}})$, where
\begin{equation}
\mu _{0}^{\mathrm{(edge)}}=(\pi ^{2}\pm |\varepsilon |)/2,  \label{E0edge}
\end{equation}
\begin{equation}
M_{\mathrm{eff}}^{\mathrm{(edge)}}=-\frac{|\varepsilon |}{2\pi
^{2}-|\varepsilon |}  \label{eff(k=pi)}
\end{equation}
and $g=(3/4)^{2}$ [recall $g$ is defined by Eq. (\ref{g})]; the signs $\pm $
in Eq. (\ref{E0edge}) pertain to two Bloch-wave solutions which differ by a
phase shift relative to the periodic OL potential in Eq. (\ref{F}). As is
seen, the effective mass in Eq. (\ref{eff(k=pi)}) is negative for the linear
excitations near the BZ edge (unless $|\varepsilon |$ is very large, in
which case the above approximation does not apply).

Equation (\ref{asympt}) with $M_{\mathrm{eff}}<0$, $g>0$, and $B<0$
(anti-trap) is tantamount to the usual 2D GP equation with positive mass,
negative $g$ (which corresponds to self-attraction), and $B>0$ (usual trap,
rather than anti-trap). As is well known \cite{rf:1,rf:2}, the GP equation
in the latter form has localized solutions, unless the norm $N$ is too large
(otherwise, collapse will take place). These solutions, and their comparison
with numerically found solutions to the full underlying equation (\ref{GP}),
will be displayed below.

An approximation similar to that based on Eq. (\ref{Bloch}) can also be
developed close to the BZ center. At the BZ center, the linear Bloch
function may be approximated by
\begin{equation}
F(x,y)=[1+c_{2}\cos (2\pi x)][1+c_{2}\cos (2\pi y)],  \label{center}
\end{equation}
and the eventual expressions for the effective mass and $\mu _{0}$ are
\begin{equation}
M_{\mathrm{eff}}^{\mathrm{(center)}}=\frac{2\pi ^{4}+\varepsilon
^{2}+\pi ^{2}\sqrt{4\pi ^{4}+2\varepsilon ^{2}}}{10\pi
^{4}+\varepsilon ^{2}-3\pi ^{2} \sqrt{4\pi ^{4}+2\varepsilon
^{2}}}~,  \label{eff(k=0)}
\end{equation}
\begin{equation}
\mu _{0}^{\mathrm{(center)}}=\pi ^{2}-\sqrt{\pi ^{4}+\varepsilon ^{2}/2}.
\label{E0center}
\end{equation}
Other coefficients are found to be $c_{2}=-\mu _{0}/\varepsilon $
and $ g=(1+12c_{2}^{2}+6c_{2}^{4})/(1+2c_{2}^{2})$. In this case,
the effective mass is always positive [in particular, in the case
with $\varepsilon =5$ and $B=0.005$, for which this approximation
will be compared to numerical results below,
$M_{\mathrm{eff}}^{\mathrm{(center)}}\approx 1.12$]. Obviously,
Eq. (\ref{eigen}) with $M_{\mathrm{eff}}>0$, $g>0$, and $B>0$ has
a localized solution (actually, of the Thomas-Fermi type, see
below) for any vorticity $m$ and norm $N$.

\section{Numerical results}

\subsection{Loosely bound states}

Our first objective is to find LB (\textquotedblleft loosely
bound\textquotedblright ) localized states near the BZ edge, which
can be constructed using the asymptotic GP equation (\ref{asympt})
and the relations (\ref{phi}) and (\ref{Bloch}), and compare them
with the corresponding numerical solutions to the full equation
(\ref{GP}). The 2D localized solutions to the asymptotic GP
equation (\ref{asympt}) were found numerically \cite{rf:1,rf:12},
and in an approximate form they can be also solved by dint of a
variational method \cite{rf:2}. In particular, the radial equation
(\ref{eigen}) can be solved numerically by the shooting method,
which simultaneously makes it possible to find the nonlinear
eigenvalue $\mu $. Figure 1 displays numerical solutions to Eq.
(\ref{eigen} ) for $m=0,1$ and $2$, obtained by means of the
shooting method. The norm of the corresponding solutions for the
underlying wave function $\phi $ is, according to Eq. (\ref{phi}),
$N\approx \int \int f^{2}(r)\cos ^{2}(\pi x)\cos ^{2}(\pi
y)dxdy\approx N_{0}/4$, where $N_{0}$ is the norm of $\Phi $ [see
Eq. (\ref{Phi})]. In particular, $N=1.52$ for $m=0$, $N=2.98$ for
$m=1$ , and $N=4.08$ for $m=2$.

\begin{figure}[tbh]
\centering\includegraphics[width=3in]{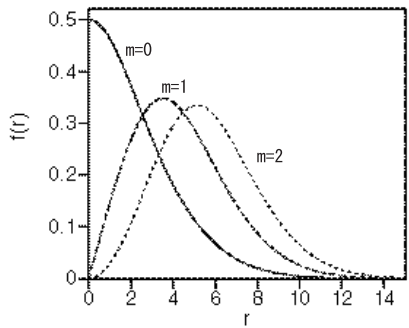} \caption{Typical
examples of numerically found solutions to Eq. (\protect\ref
{eigen}) with $M_{\mathrm{eff}}=-10/(2\protect\pi^2-10),
g=(3/4)^2$ and $ B=-0.005$ which represent 2D solitons with
vorticity $m=0,1$ and $2$.} \label{fig1}
\end{figure}

To characterize families of the solutions, Figs. 2(a) and 2(b)
display, respectively, the absolute value of the energy-per-atom,
$-\epsilon =-E/N$ for the solution [the energy $E$ is given by the
functional (7)], and the maximum value of $f(r)$ vs. the norm $N$.
In particular, $-\epsilon \ $ decreases with $N$, because a larger
atom density gives rise a larger positive contribution to $E$
through the term $g|\Phi |^{4}$ in the energy density. The maximum
value of $f(r)$ increases with $N$, since the effective
interaction is attractive owing to the negativeness of
$M_{\mathrm{eff}}$.

\begin{figure}[tbh]
\centering\includegraphics[width=4in]{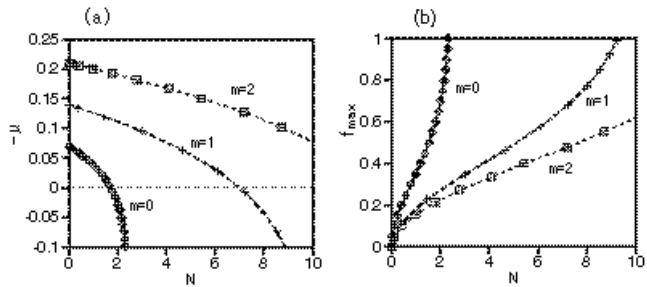}
\caption{The absolute value of the energy-per-atom, $-\protect\epsilon $,
(a) and amplitude (b) of the 2D solitons with the vorticity $0,1$ and $2$,
as found from Eq. (\protect\ref{eigen}), vs. the norm.}
\label{fig2}
\end{figure}

Figures 3(a), (b) and (c) present comparison of the above
semi-analytical solutions (for $m=0,1$ and $2$), based on Eqs.
(\ref{phi}), (\ref{Bloch}) and (\ref{asympt}), with direct
numerical solutions of the full 2D equation ( \ref{GP}). The
comparison is performed by displaying the semi-analytical and
direct numerical solutions for $|\phi (x,L/2)|$, taken along the
central cross-section, at $y=L/2$ (the norms of the numerical and
semi-analytical solutions are virtually identical in each case).
Fairly good agreement is seen for $m=0,1$ and $2$, although $|\phi
|$ in the direct numerical solutions is slightly larger.
\begin{figure}[tbh]
\centering\includegraphics[width=5in]{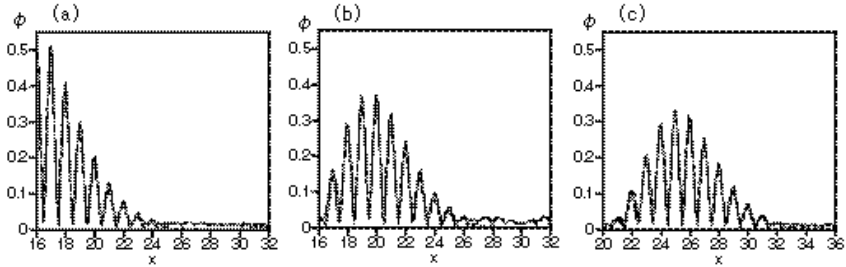}
\caption{The central cross section of solutions $|\protect\phi |$ for the
loosely bound 2D states near the edge of the Brillouin zone with (a) $m=0$,
(b) $m=1$, and (c) $m=2$, which corresponds to Eq. (\protect\ref{Bloch}).
Solid and dashed curves depict, respectively, the solutions obtained from
the asymptotic equation (\protect\ref{asympt}) and full GP equation (\protect
\ref{GP}).}
\label{fig3}
\end{figure}

\begin{figure}[tbh]
\centering\includegraphics[width=5in]{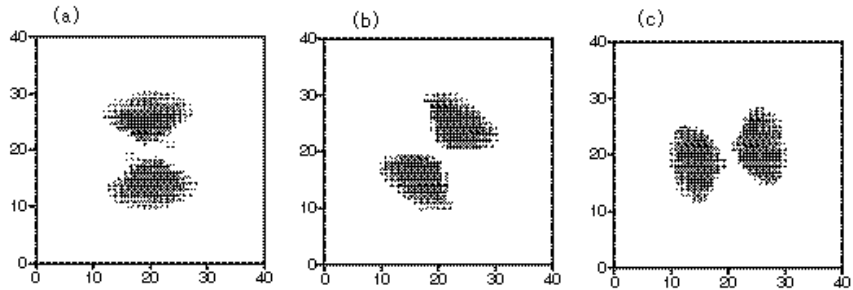}
\caption{An example of the uniform rotation of the vortex state with $N=4.08$
and $m=2$.}
\label{fig4}
\end{figure}

From the viewpoint of the reduced GP equation (\ref{asympt}), which does not
explicitly contain the OL potential, the angular velocity $\Omega =\mu /m$,
which appears in the expression (\ref{Phi}), is only a phase velocity.
However, the relation (\ref{phi}) suggests that the corresponding solution
to the full GP equation (\ref{GP}) may feature \emph{actual rotation} of the
pattern with the angular velocity $\Omega $. Direct simulations confirm this
expectation, as shown in Fig. \ref{fig4}. The figure displays three
consecutive snapshots of the region in which the vortex solution with $m=2$
shown in Fig. 3(c) takes sufficiently large values, namely, satisfying
conditions $\left[ \cos (\pi x)\cos (\pi y)\right] \mathrm{Re}\phi (x,y)>0$
and $|\phi (x,y)|~>0.06$. The time interval between the snapshots is $1$.
This figure clearly shows that the vortex state indeed rotates with the
frequency $\Omega \approx \mu /m$ in the OL potential. This figure also
demonstrates that the vortex state with $m=2$ is \emph{stable}. The vortex
solutions, with $m=1$ and $2$, to the asymptotic equation (\ref{asympt}) are
known to exist \cite{rf:12} and be dynamically stable \cite{rf:21} if $N$ is
smaller than a critical value. We have also checked the stability of the
vortex states with $m=1$ and $2$, shown in Fig.~3, by direct numerical
simulations of Eq.~(2), starting from initial conditions in the form of
perturbed vortex states. When the norm $N$ is increased, more complicated behaviors are expected. Some simulation results such as a transition to a tightly bound state and a symmetry-breaking process are shown in the next section.

\begin{figure}[tbh]
\centering\includegraphics[width=5in]{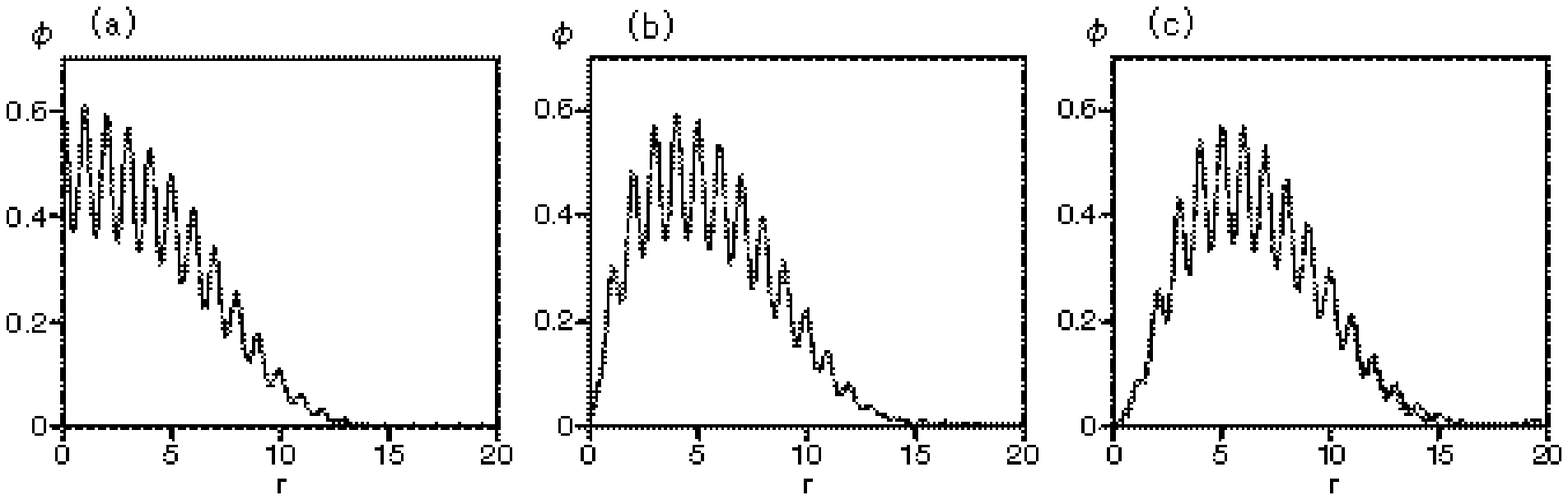} \caption{Dashed
curves correspond to Thomas-Fermi-like solutions of the asymptotic
equation (\protect\ref{asympt}) with $B=+0.005$ (the usual trap,
rather than anti-trap) and positive effective mass,
$M_{\mathrm{eff}}=1.12$, as found by the shooting method from
Eq.~(9). These solutions describe loosely bound states close to
the center of the Brillouin zone, as per Eq. (
\protect\ref{center}). Continuous curves are numerical
counterparts of these solutions, obtained from the full GP
equation ( \protect\ref{GP}) with $ \protect\varepsilon =5$. (a)
$m=0$, $N=19.7$; (b) $m=1$, $N=29.8$; (c) $m=2$ , $N=33.0$.}
\label{fig5}
\end{figure}

Solutions predicted by the asymptotic equation (\ref{asympt}) near the BZ
center, i.e., corresponding to Eqs. (\ref{center}) and (\ref{eff(k=0)}),
were also found, and compared to direct numerical solutions of the
underlying GP equation (\ref{GP}). It should be stressed that, as the
effective mass (\ref{eff(k=0)}) is positive in this case, the corresponding
localized solutions to the asymptotic equation (\ref{asympt}) with $g>0$
(repulsive interaction) are not soliton-like ones, but are instead close to
the Thomas-Fermi (TF) states, that can be obtained neglecting the
kinetic-energy term in the GP equation \cite{Pit}. Figure \ref{fig5} shows
typical examples of thus found solutions for $m=0$, $m=1$, and $m=2$. As is
seen, the accuracy provided by Eqs. (\ref{asympt}) and (\ref{center}) is
very good in this case, for all the values of $m$.

\subsection{Relation to tightly bound solitons}

\begin{figure}[tbh]
\centering\includegraphics[width=4in]{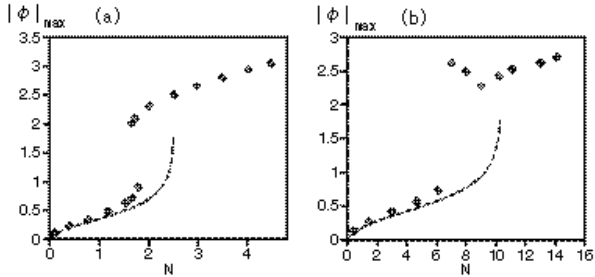}
\caption{The amplitude of the 2D solution with $m=0$ (a) and $m=1$ (b) vs.
the norm (number of atoms) $N$, in the case of $\protect\varepsilon =10$ and
$B=-0.005$ [the dashed lines show the same dependencies, as predicted by the
asymptotic equation (\protect\ref{asympt}) for the loosely bound states].
The lower and upper branches of the dependence correspond to the loosely and
tightly bound states, respectively (see further details in the text).}
\label{fig6}
\end{figure}

The full GP equation (\ref{GP}) cannot be reduced to the
asymptotic NLS equation (\ref{asympt}) if the assumption of the
slow variation of the envelope function is not valid; in that
case, strongly confined (\textquotedblleft tightly
bound\textquotedblright , TB) states may be possible, that are not
described by Eq. (\ref{asympt}). To investigate such solutions,
and compare them to their loosely bound counterparts, we performed
direct numerical integration of Eq.~(\ref{GP}), looking for
solutions that could be classified as having the vorticity $m=0$
or $m=1$ [in terms of Eq. (\ref{Phi})]. Figures \ref{fig6}(a) and
\ref{fig6}(b) display the amplitude of the thus found solutions
[the maximum of $|\phi (x,y)|$] as a function of the norm, $N=\int
\int |\phi |^{2}dxdy$. The lower branches in these figures
precisely correspond to the LB states described above, the dashed
curves showing the dependence of the amplitude vs. $N$ as found
from the asymptotic equation (\ref{asympt}). The solutions which
belong to the upper branches in Fig.~6 can be identified as TB
(strongly localized)\ 2D gap-soliton states in the GP equation
with repulsive interaction between atoms, that persist in the
limit of $B=0$ and were first found in Refs. \cite{SalernoJPhysB}.
There is weak
hysteresis for $m=0$ between the LB and TB states for the parameter values of $\epsilon=10$ and $B=-0.005$. The discontinuous transition between the LB and TB states occurs for sufficiently large $\epsilon$. (The discontinuous transition was not observed for $\epsilon$=5 and $B=-0.005$.)
 Figures 6(a) and (b) also make it obvious that the asymptotic equation
(\ref{asympt}) predicts a jump upward from the lower branch, which
corresponds to the collapse in the asymptotic 2D NLS equation
(\ref{asympt}). However, the collapse does not occur in the full
GP equation (\ref{GP}) with the repulsive nonlinearity (note
that the negative effective mass is irrelevant when the
localization length is on the same order of magnitude as the OL
period). The critical value of the norm for the collapse in the
asymptotic equation (\ref{asympt}) is somewhat larger than the
value at which the LB-TB transition occurs in the numerical
solution of Eq. (\ref{GP}).

It is necessary to estimate the actual number of atoms in the BEC
at which the LB-TB transition, shown in Fig.~6, is expected.
Assuming a \textquotedblleft pizza-shaped\textquotedblright\
condensate, with the diameter $\sim 1000$ OL periods and
transverse width ten times smaller, taking the scattering length
$5$ nm (which is close to its value for $^{87}$ Rb), and following
once again the derivation of the 2D equation (\ref{GP}) from the
full 3D GP equation, we conclude that the real number of atoms is
obtained by multiplying the numbers on the horizontal axes in
Fig.~6 by a factor $\simeq 10^{3}$.

To further illustrate the drastic difference between the LB and TB
states with zero vorticity ($m=0$), we display their typical
examples, found as direct numerical solutions of Eq. (\ref{GP}),
in Figs.~7(a) and (b). These figures display regions in which the
absolute value of the respective solution exceeds the
quarter-amplitude, i.e., $|\phi (x,y)|~>|\phi |_{
\mathrm{max}}/4$. Additionally, Fig. 7(c) displays the absolute
value for both solutions, $|\phi (x,x)|$, as a function of $x$
along the diagonal, $ x=y $.

\begin{figure}[tbh]
\centering\includegraphics[width=5in]{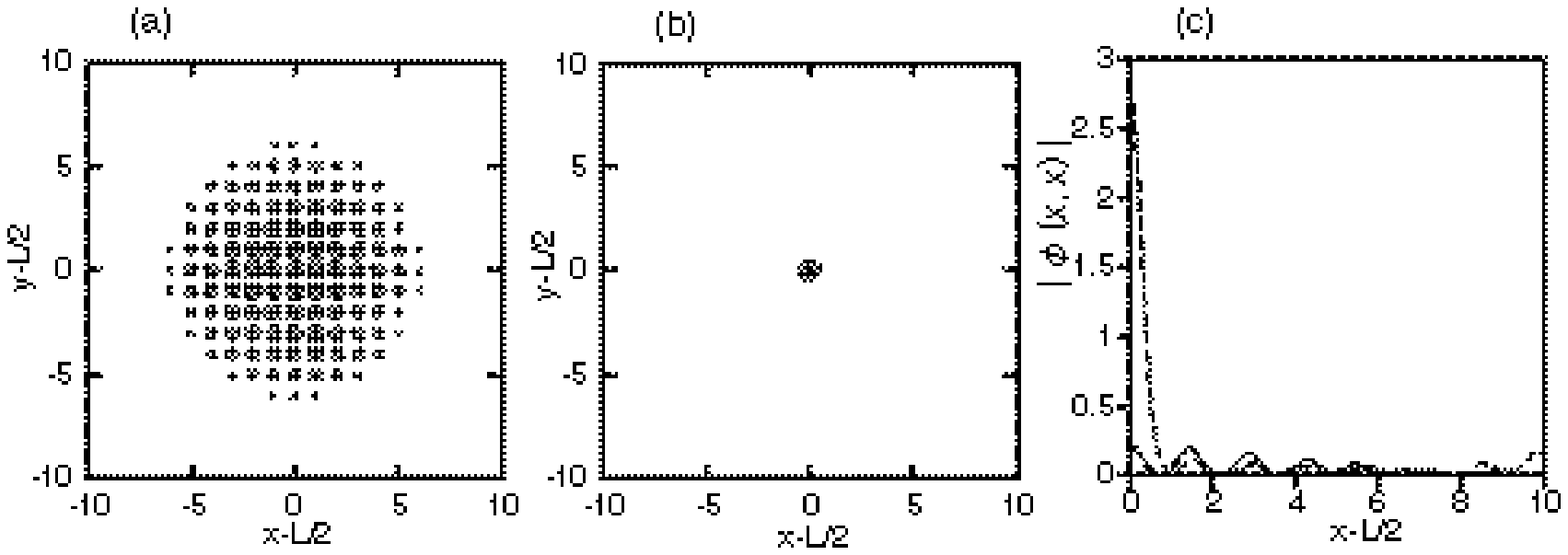}
\caption{Typical examples of the loosely (a) and tightly (b) bound 2D states
with $m=0$ (zero vorticity) for $\protect\varepsilon =10$ and $B=-0.005$.
The norms of the two states are, respectively, $N_{\mathrm{loose}}=0.393$
and $N_{\mathrm{tight}}=2.66$. Panel (c) displays the diagonal cross
sections of the two solutions, $\left\vert \protect\phi (x,x)\right\vert $
(the solid and dashed lines depict the loosely and tightly bound states,
respectively).}
\label{fig7}
\end{figure}
\begin{figure}[tbh]
\centering\includegraphics[width=5in]{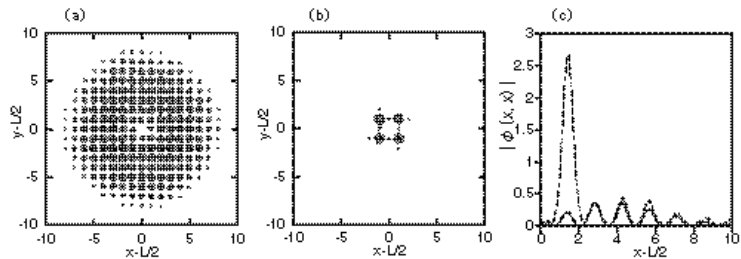}
\caption{The same as in Fig. \protect\ref{fig7} for the loosely (a) and
tightly (b) bound vortices with $m=1$. The norms of the two states are,
respectively, $N_{\mathrm{loose}}=0.413$ and $N_{\mathrm{tight}}=13$.}
\label{fig8}
\end{figure}

\begin{figure}[tbh]
\centering\includegraphics[width=4in]{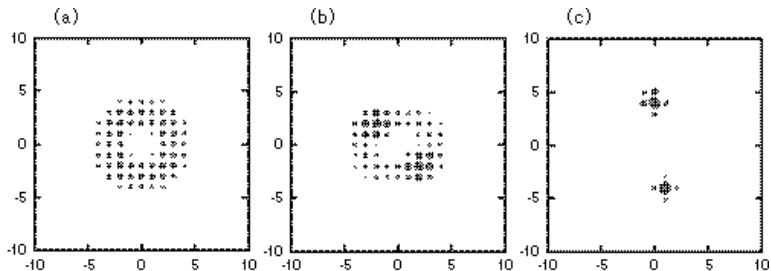} \caption{Three
snapshot patterns taken at the moments (a) $t=10$, (b) $t=40$, and
(c) $t=80$ in the simulation of a loosely-bound vortex with $m=1$,
evolving into a tightly-bound state. The norm is $N=7$.}
\label{fig9}
\end{figure}

\begin{figure}[tbh]
\centering\includegraphics[width=5in]{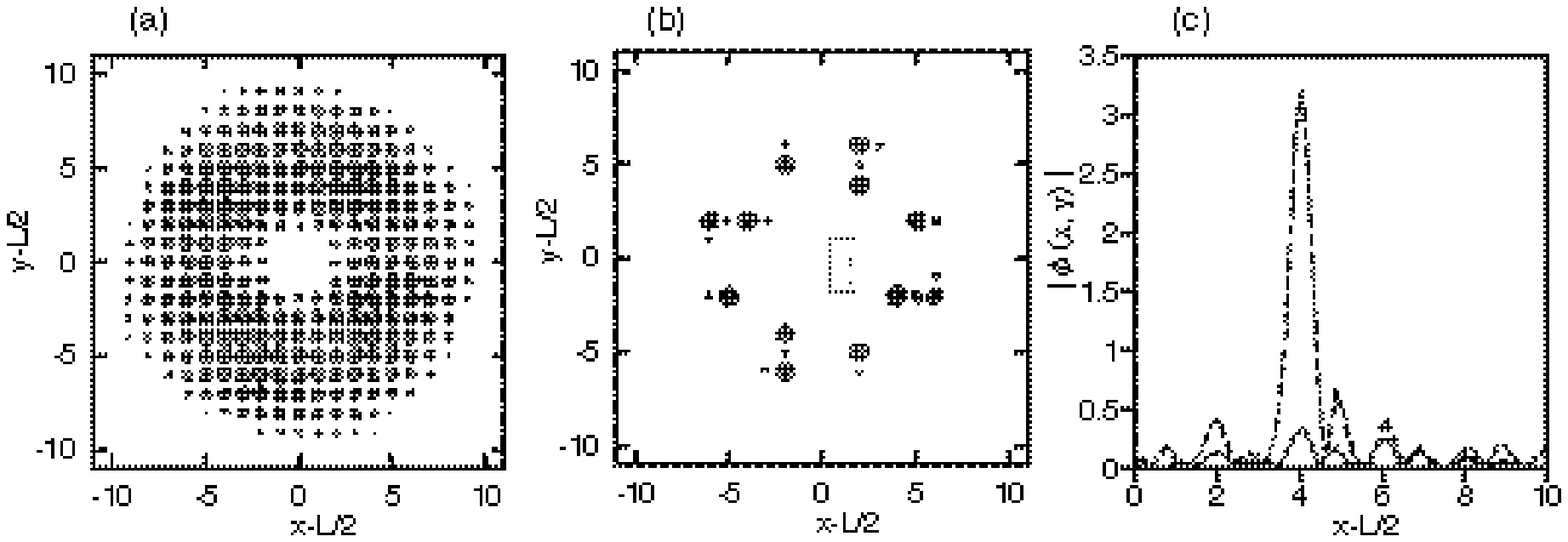} \caption{The same
as in Figs. \ref{fig7} and \ref{fig8} for the loosely (a) and
tightly (b) bound vortices with $m=2$. The norms of the two states
are, respectively, $N_{\mathrm{loose}}=3.92$ and
$N_{\mathrm{tight}}=48.1$. (c) A profile of $|\protect\phi (x,y)|$
for the tightly-bound vortex along the line $y-L/2=-0.45(x-L/2)$.}
\label{fig10}
\end{figure}

In a similar manner, the comparison of the LB and TB states with
$m=1$ is presented in Fig.~8, where the panels (a) and (b) again
display regions where, respectively, the loose and tight solutions
take the values with $ |\phi (x,y)|~>|\phi |_{\mathrm{max}}/4$,
and the panel (c) shows the diagonal cross sections of both
solutions. In particular, the TB vortex in Fig.~8(b) features a
four-fold symmetry (it is invariant with respect to the rotation
by $\pi /2$). In this connection, it is relevant to mention that,
while the same symmetry is observed in all the TB vortices for a
sufficiently large norm, $N\geq 10$, the four-fold symmetry is
broken for $N\leq 9$.

To illustrate the latter property, Fig. 9 displays three
consecutive snapshots of regions with $|\phi|>0.5$, taken from the
simulation starting with an LB vortex state that has $m=1$ and
$N=7$. As is seen, the patterns do not feature the four-fold
symmetry, unlike Fig.~8(b), but a nearly two-fold symmetric TB
state is generated.

The transition between the less symmetric and more symmetric TB
vortices accounts for the slope breaking in the amplitude-vs.-norm
curve for $N=9$, which is evident in Fig.~6(b). We have also found
the TB\ vortices with $m=2$ and checked that they are stable as
well. Figure 10 displays the TB vortices with $m=2$ for $N=3.92$
and $N=48.1$. Figures 10(a) and (b) display the regions where the
TB solutions take the values with $|\phi (x,y)|~>|\phi
|_{\mathrm{max}}/4$, and the panel (c) shows the cross sections
along the line $y-L/2=-0.45(x-L/2)$, on which $|\phi |$ takes the
maximum value for $N=48.1$. To the best of our knowledge, the
existence of TB vortex solitons in the 2D model with
self-repulsion has not been reported before, therefore theses
result (displayed here for $m=1$ and $2$) are novel by themselves
too (similar stable vortices were very recently found by means of
a different method \cite{BMS2}).

The localized solutions belonging to the LB and TB branches differ
drastically not only in the dependence of $|\phi |_{\mathrm{max}}$
vs. $N$ (i.e., in their static properties), but in dynamical
properties too. As it will be shown in the next section, the TB
solitons are always pinned by the underlying OL potential, while
the LB states may move freely. In fact, a sharp transition between
mobile and immobile 1D solitons with $M_{\mathrm{eff }}<0$ (i.e.,
close to the BZ edge) and $g>0$ (repulsive interactions) was
reported in Ref. \cite{rf:5}. As well as in the 2D case, the
immobilization takes place with the increase of the soliton's
norm, and no hysteresis between the mobile and immobile solitons
is found.

The abrupt transition between the LB and TB localized states, as
described above, is a characteristic feature of the localized
states with the negative effective mass, found close to the BZ
edge, where the LB state is well approximated by Eqs. (\ref{phi}),
(\ref{Phi}), and (\ref{Bloch}).\ On the other hand, no similar
transition is found for the localized states close to the BZ
center, where the approximation (\ref{Bloch}) is replaced by (\ref
{center}). The absence of the distinction between LB and TB states
in the latter case is explained by the fact that, as is commonly
known, TB solitons simply do not exist in the case of the
self-repulsion ($g>0$) near the BZ center. Localized LB states of
course exist, as suggested by the TF approximation.

\section{Motion of loosely bound localized states}

In the 2D model with both repulsive \cite{SalernoJPhysB} and
attractive \cite {BMS} interaction, the TB solitons are strongly
pinned by the OL potential, and cannot move (stable TB solitons
also exist in the 2D attractive model with a quasi-1D periodic
potential, and in the 3D model with a quasi-2D potential, in which
case they may freely move in the unconfined direction
\cite{BMS2}). On the contrary to that, the LB localized states
described above may move freely on top of the underlying 2D
periodic potential, keeping their coherent structure. This feature
resembles a similar one that we have recently demonstrated in the
1D version of the present model \cite {rf:5}; in particular, the
1D solitons with the negative effective mass are expelled by the
ordinary parabolic trap, and are held by the anti-trap. However,
the character of the motion in the 2D case may be completely
different, see below.

A 2D LB state can be set in motion either by an initial shift from
the central position [relative to the (anti-)trap], or by lending
it an initial momentum. We have performed numerical simulations,
using both ways to push the localized state. For instance, Fig.~10
displays results obtained with the initial condition $\phi
(x,y,0)=\phi _{0}(x-5,y)\exp \left( ikq_{y}y\right) $, where $\phi
_{0}(x,y)$ is the stationary LB state with $ m=0$ and $N=0.479$,
whose peak position is located at the central point, $ (L/2,L/2)$,
and $q_{y}=0.2$. In other words, this initial condition implies
that the localized state is pushed by shifting its center in the
$x$ direction, and giving it an initial momentum along $y$.

We define the instantaneous position of the center of the
localized state as $\mathbf{X}=\int \int \mathbf{x}|\phi
(x,y;t)|^{2}dxdy/\int \int |\phi |^{2}dxdy$. Figure 11(a)
demonstrates that, as a result, the center of the LB state moves
along an elliptic trajectory [in the case displayed in Fig.~11(a),
the simulations were run up to the time $t_{\max }=100$, while the
numerically measured period of the elliptic motion was
$T_{\mathrm{num} }\approx 88$, i.e., the localized state has
completed one full circulation (in the clockwise direction, since
the initial velocity $v_{y}=q_{y}/M_{ \mathrm{eff}}$ is negative).
To make the difference from the strongly pinned TB solitons
clearer, Fig.~11(b) displays a result of a similar attempt to set
in motion the localized state of the TB type. As a result,
initially its center moves irregularly, and is then trapped by the
OL potential. Many other runs of the simulations produced quite
similar results; in particular, the eccentricity of the elliptic
orbit depends on the initial push, and effectively 1D periodic
motion along straight lines through the center of the domain are
possible too. In this connection, it is relevant to mention that,
although the underlying OL makes the medium anisotropic,
simulations do not reveal any dependence of the frequency on the
direction of the 1D oscillations (the frequency of the elliptic
motion does not depend on the orientation of the ellipse either).

\begin{figure}[tbh]
\centering\includegraphics[width=4in]{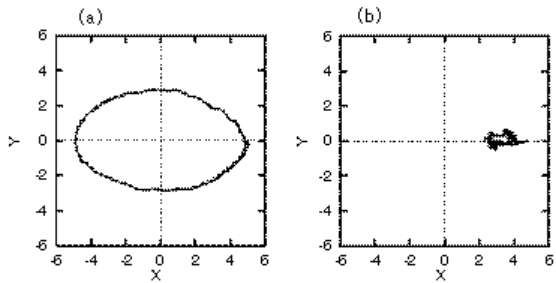}
\caption{Panel (a) displays a typical example of the stable elliptic motion
of a localized loosely-bound state with zero vorticity ($m=0$) and negative
effective mass ($M_{\mathrm{eff}}\approx -1.03$) in the \emph{anti-trap}
with $B=-0.005$. In this case, $\protect\varepsilon =10$, and the norm of
the localized state is $N=0.479$. For comparison, panel (b) displays an
attempt to set in motion a tightly-bound soliton with $N=4.01$.}
\label{fig11}
\end{figure}

The periodic motion of the LB state can be easily described within the
framework of the asymptotic equation (\ref{asympt}). In that case, one can
demonstrate that motion for the center $\mathbf{X}$ of the localized state,
which is treated as a rigid quasi-particle, obeys the Newton's equation of
motion,
\begin{equation}
M_{\mathrm{eff}} \frac{d^{2}\mathbf{X}}{dt^{2}}=-B\mathbf{X}
\label{Newton}
\end{equation}
[$M_{\mathrm{eff}}$ is the same effective mass as in Eq.
(\ref{asympt}), while the right-hand side is the potential force
generated by the parabolic (anti-)trap]. Equation (\ref{Newton})
indeed describes the observed motion with a good accuracy. For
instance, in the case displayed in Fig. \ref{fig8} (a), Eq.
(\ref{Newton}) predicts the period of the motion,
\begin{equation}
T_{\mathrm{anal}}=2\pi \sqrt{M_{\mathrm{eff}}/B}\approx 90,  \label{T}
\end{equation}
which should be compared to the above-mentioned numerically found
period, $ T_{\mathrm{num}}\approx 88$.$\allowbreak $

As it was explained above, the localized states found close to the BZ center
are actually always of the LB type, hence one may expect that they may move
freely too. This is indeed observed in simulations. Typical examples are
displayed in Fig.~12, for the localized states with $m=0$ and $2$ (quite a
similar motion mode was observed for the vortex with $m=1$). The trajectory
corresponds to anti-clockwise elliptic circulation in this case, as the
initial velocity $v_{y}=q_{y}/M_{\mathrm{eff}}$ is positive.

We stress that the vortex LB states readily exhibit circular motion with the
same period as their counterparts with $m=0$, and the analytical prediction
for the circulation period, based on Eq. (\ref{T}), is very accurate in this
case too. For example, in both cases ($m=0$ and $m=2$) shown in Fig.~12, the
predicted period is $T_{\mathrm{anal}}\approx 94.0$, while the one extracted
from the numerical data is $T_{\mathrm{num}}\approx 94.5$.

\begin{figure}[tbh]
\centering\includegraphics[width=5in]{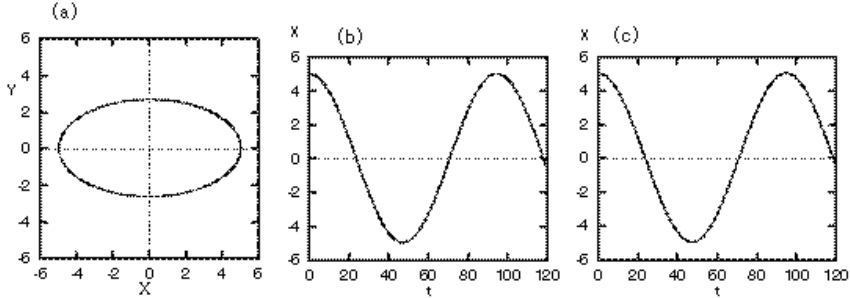}
\caption{Panel (a) has the same meaning as in Fig.~11, but for the case of
the positive effective mass, $M_{\mathrm{eff}}\approx 1.12$, and in the
ordinary trap with $B=+0.005$ (and $\protect\varepsilon =5$). Panel (b)
shows the corresponding time dependence of the central coordinate $X$ of the
moving localized state. In panel (c), the same dependence is shown for the
circular motion of the vortex with $m=2$.}
\label{fig12}
\end{figure}

\section{Conclusion}

In this work, we have developed analysis of dynamics of nonlinear
localized excitations in two-dimensional (2D) Bose-Einstein
condensates (BECs) with repulsive interactions, which are
subjected to the action of the optical lattice (OL) combined with
the external parabolic potential. First, we have shown, in an
analytical form, that broad (\textquotedblleft loosely
bound\textquotedblright , LB) localized states, based on the 2D
Bloch function near the edge of the Brillouin zone (BZ), feature a
negative effective mass (while the mass of the LB localized states
is positive near the BZ center). The negative-mass pulse cannot be
confined by the usual trap, but it is held by an inverted
parabolic potential (\textit{anti-trap} ). Direct simulations have
demonstrated that the LB states (including vortices) are stable
and can freely move on top of the OL, and the frequency of their
periodic motion along an elliptic orbit in the anti-trapping
potential agrees very well with the analytical prediction, which
treats the LB solitons as quasi-particles. Besides that, the LB
vortices feature rotation around the center. With the increase of
the number of atoms, an abrupt transition occurs, from the
negative-mass LB states to their tightly bound (TB) counterparts;
unlike the LB states, the TB solitons are rigidly pinned by the
OL. The positive-mass LB states (including vortices), constructed
near the BZ center, are stable and move freely too.

Further dynamical effects may be considered in the 2D situation, such as
collisions between moving solitons. In fact, various types of collisions are
possible, such as between zero-vorticity solitons and vortices, as well as
between LB and TB states with the negative mass (the latter is possible in
the presence of the anti-trap). On the other hand, interactions between
solitons with positive and negative mass are precluded, as only one type of
the solitons may be confined by a given trapping or anti-trapping potential.
The consideration of collisions is, however, beyond the scope of this paper.

We expect that similar LB and TB states may be found in BECs subject to the
action of the 3D (three-dimensional) OL potential. However, simulations of
the corresponding 3D GP equation is a rather difficult problem.

Estimates for physical parameters at which the predicted effects
may be observed are similar to those in the 1D counterpart of the
model \cite{rf:5} . In particular, a circular motion of the LB
soliton in the anti-trap may be expected at a frequency $\sim 100$
Hz in a domain of the diameter $\sim 0.5$ mm ($\simeq 1000$
spatial periods of the OL). The transition from the LB to TB
solitons is expected when the number of atoms in the condensate
attains the values between $\simeq 2\cdot 10^{3}$ for the $m=0$
states and $\simeq 6\cdot 10^{3}$ for the vortex with $m=1$.

In the experiment, the LB\ and TB localized states are expected to form
spontaneously from a properly chosen number of atoms loaded in the
appropriate potential, as there is virtually no overlap between these two
types of the states. As concerns vortex states, they may be created using
the well-known technique of optical stirring. The effects predicted in this
work for the BEC can also take place with spatial optical solitons in bulk
photonic crystals.

\section*{Acknowledgements}

The work of B.A.M. was supported, in a part, by the Israel Science
Foundation through a grant No. 8006/03.

\end{document}